\documentclass[preprint,12pt, a4paper]{elsarticle}

\usepackage{amssymb}
\usepackage{hyperref}
\usepackage{subcaption}
\usepackage{listings}
\usepackage{multirow}
\usepackage{threeparttable}

\journal{SoftwareX}

\begin{document}
\renewcommand{\labelenumii}{\arabic{enumi}.\arabic{enumii}}

\begin{frontmatter}



\title{ComputeFHE: A Privacy-Preserving General-Purpose Computation Library}

\author[author1]{Faris Serdar TAŞEL}
\author[author2]{Efe ÇİFTCİ}
\address[author1]{Department of Computer Engineering, Çankaya University, Ankara, Turkey, fst@cankaya.edu.tr}
\address[author2]{Computer Programming Program, Çankaya University, Ankara, Turkey, efeciftci@cankaya.edu.tr}


\begin{abstract}
Fully Homomorphic Encryption (FHE) enables computations to be performed directly on encrypted data while preserving data confidentiality. However, its practical applications remain limited by high computational costs and development complexity. This paper presents ComputeFHE, an open-source C++ library that facilitates the development of privacy-preserving applications based on the TFHE cryptosystem. The library provides encrypted integer and fixed-point data types together with arithmetic, logical, comparison, conditional, and oblivious array-access operations which allow developers to implement algorithms using a familiar imperative programming paradigm. ComputeFHE supports both conventional TFHE arithmetic based on standard two-input logic gates and an optimized Arithmetic Logic Unit (ALU) architecture utilizing FHE-friendly logic primitives. Experimental results demonstrate significant reductions in the number of required bootstrapping operations, achieving performance improvements of up to 3.9× for selected operations. In addition, the library includes a simulation mode that enables testing, debugging, and complexity analysis without performing actual cryptographic computations while providing circuit complexity and bootstrapping costs. Built on top of OpenFHE, ComputeFHE offers a practical and accessible framework for developing and evaluating privacy-preserving algorithms and applications.
\end{abstract}


\end{frontmatter}


\section*{Required Metadata}
\label{}

\section*{Current code version}
\label{}


\begin{table}[!ht]
\begin{tabular}{|l|p{6.5cm}|p{6.5cm}|}
\hline
\textbf{Nr.} & \textbf{Code metadata description} & \textbf{Metadata} \\
\hline
C1 & Current code version & 1.0 \\
\hline
C2 & Permanent GitHub link to code/repository used for this code version & \url{https://github.com/fstasel/compute-fhe} \\
\hline
C3 & Legal Code License   & MIT \\
\hline
C4 & Code versioning system used & Git \\
\hline
C5 & Software code languages, tools, and services used & C++, CMake. \\
\hline
C6 & Compilation requirements, operating environments \& dependencies & A modern GNU/Linux system with development files for OpenFHE installed. Alternatively, a Dockerfile is provided for easily setting up a preconfigured environment.\\
\hline
C7 & If available Link to developer documentation/manual & \url{https://github.com/fstasel/compute-fhe#readme} \\
\hline
C8 & Support email for questions & fst@cankaya.edu.tr \\
\hline
\end{tabular}
\caption{Code metadata}
\label{} 
\end{table}


\begin{enumerate}

\item Motivation and significance\\

Homomorphic encryption is an encryption paradigm that enables computations to be performed directly on encrypted data without requiring the underlying plaintext to be decrypted \cite{doan2023survey}. This encryption technique allows privacy-sensitive information shared in encrypted form to be processed by a service provider while ensuring that the computation results also remain encrypted. When necessary, the resulting ciphertext can later be decrypted using the appropriate key to reveal the original content.\\

The variant of homomorphic encryption that supports an unrestricted number of computations on encrypted data is referred to as Fully Homomorphic Encryption (FHE) \cite{doan2023survey,gentry2009fully}. Prominent examples include TFHE (Torus FHE, also known as CGGI--Chillotti-Gama-Georgieva-Izabachène) \cite{chillotti2016faster,chillotti2017faster} and CKKS (Cheon-Kim-Kim-Song) \cite{cheon2018full}.\\

In FHE schemes, one of the key elements ensuring data security is noise\footnote{\url{https://docs.zama.org/tfhe-rs/explanations/tfhe-deep-dive}}. During homomorphic computations, this noise accumulates progressively, eventually rendering further computations infeasible. Therefore, to enable continuous processing of encrypted data, a procedure known as bootstrapping must be periodically applied to reduce the noise level \cite{bai2024research}. However, in current approaches, bootstrapping requires substantial computational resources and constitutes a major bottleneck in practical applications \cite{bergerat2025accelerating,xiang2023fast,chillotti2021improved}. Consequently, any optimization capable of improving efficiency is of critical importance in FHE systems.\\

TFHE is a prominent cryptosystem that enables unlimited encrypted computation through basic logic gates or lookup tables (LUTs) \cite{chillotti2016faster,chillotti2017faster}. In this cryptosystem, the complexity of the bootstrapping operation depends on both the complexity of the homomorphic computation being performed and the targeted security level.\\

ComputeFHE is a library designed for general-purpose computation based on the TFHE gate-based approach, utilizing optimized gates that reduce the number of bootstrapping operations and enable encrypted-data processing using C-style programming techniques. Through this library, developers can use encrypted data for various purposes, including arithmetic, logical and comparison operations, conditional execution, and oblivious array access\footnote{Reading or writing an array element without revealing which index was accessed, hiding the memory access pattern from observers.}, which allows for the implementation of privacy-preserving variants of diverse algorithms.\\

OpenFHE\footnote{\url{https://github.com/openfheorg/openfhe-development}} library, which is used by ComputeFHE as its backend, provides support for fundamental homomorphic operations across various FHE cryptosystems (e.g., TFHE and CKKS) and multiple security levels. Another important library capable of performing tasks similar to those of ComputeFHE is Concrete\footnote{\url{https://github.com/zama-ai/concrete}}, which enables privacy-preserving computations by leveraging TFHE’s non-linear lookup table (LUT) mapping capability \cite{Concrete}. In addition, ConcreteML\footnote{\url{https://github.com/zama-ai/concrete-ml}}, built on top of Concrete, applies machine learning methods within the homomorphic encryption framework \cite{ConcreteML}. Furthermore, TFHE also has a GPU-accelerated Rust implementation called TFHE-rs\footnote{\url{https://github.com/zama-ai/tfhe-rs}} which provides low-level backend \cite{TFHE-rs}.\\

HELM\footnote{\url{https://github.com/TrustworthyComputing/helm}} is a compiler-based toolchain designed to advance privacy-preserving computation by automatically converting high-level hardware descriptions into efficient, FHE-compatible circuits. It maps complex logic into a variety of gate-level or lookup table structures based on the target circuit's specific requirements \cite{gouert2025helm}. Furthermore, BOLT introduces a specialized synthesis layer that focuses on the high computational cost of bootstrapping. It rewrites sub-networks into more compact, functionally equivalent forms to reduce the necessary number of bootstrapping operations \cite{chaturvedi2026bolt}.\\

\item Software description\\

ComputeFHE is a C++ library based on the TFHE cryptosystem that performs homomorphic arithmetic and logical operations on encrypted data types using logic-gate components. The library enables algorithmic design on encrypted data without departing from the principles of imperative programming. In this way, developers who are not experts in homomorphic computation processes can still contribute to the development of algorithms that allow service providers to process data in a privacy-preserving manner, while remaining within the familiar programming paradigm and syntax of conventional C++. To facilitate computation over encrypted data, ComputeFHE provides predefined encrypted integer and fixed-point data types, as well as encrypted conditional statements and vector structures that are fully compatible with these data types.\\

ComputeFHE provides both standard implementations of Arithmetic Logic Unit (ALU) operations using basic logic gates and optimized implementations based on the FHE-friendly specialized gates proposed in \cite{tasel2025improved}. In addition, the ComputeFHE library enables simulation of the developed application without requiring actual high-performance computational resources. It can also generate statistical metrics, such as the number of bootstrapping operations and logic-gate utilization, to estimate the actual computational cost.\\

ComputeFHE utilizes the OpenFHE library to provide the fundamental TFHE functionalities, including key generation, encryption, decryption, and homomorphic computation. The OpenFHE library employs recommended cryptosystem parameters which corresponds to various security levels, such as 128-, 192-, and 256-bit security, supports multiple bootstrapping techniques, and provides implementations of fundamental logic-gate operations, including AND, OR, XOR, and related binary operators. Furthermore, OpenFHE supports HEXL \cite{IntelHEXL} library and leverages AVX-512 processor instructions, and ComputeFHE is naturally compatible with it as well.\\
    
\begin{enumerate}
    \item Software architecture:
    The ComputeFHE library consists of the core, C++ utilities, and simulator modules to perform homomorphic arithmetic and logical operations. A class diagram of the library is illustrated in Figure~\ref{fig:uml}.\\

    \begin{figure}[bt]
    {\phantomsubcaption\label{fig:uml}\phantomsubcaption\label{fig:uml}}
    \centering
    \includegraphics[width = \linewidth]{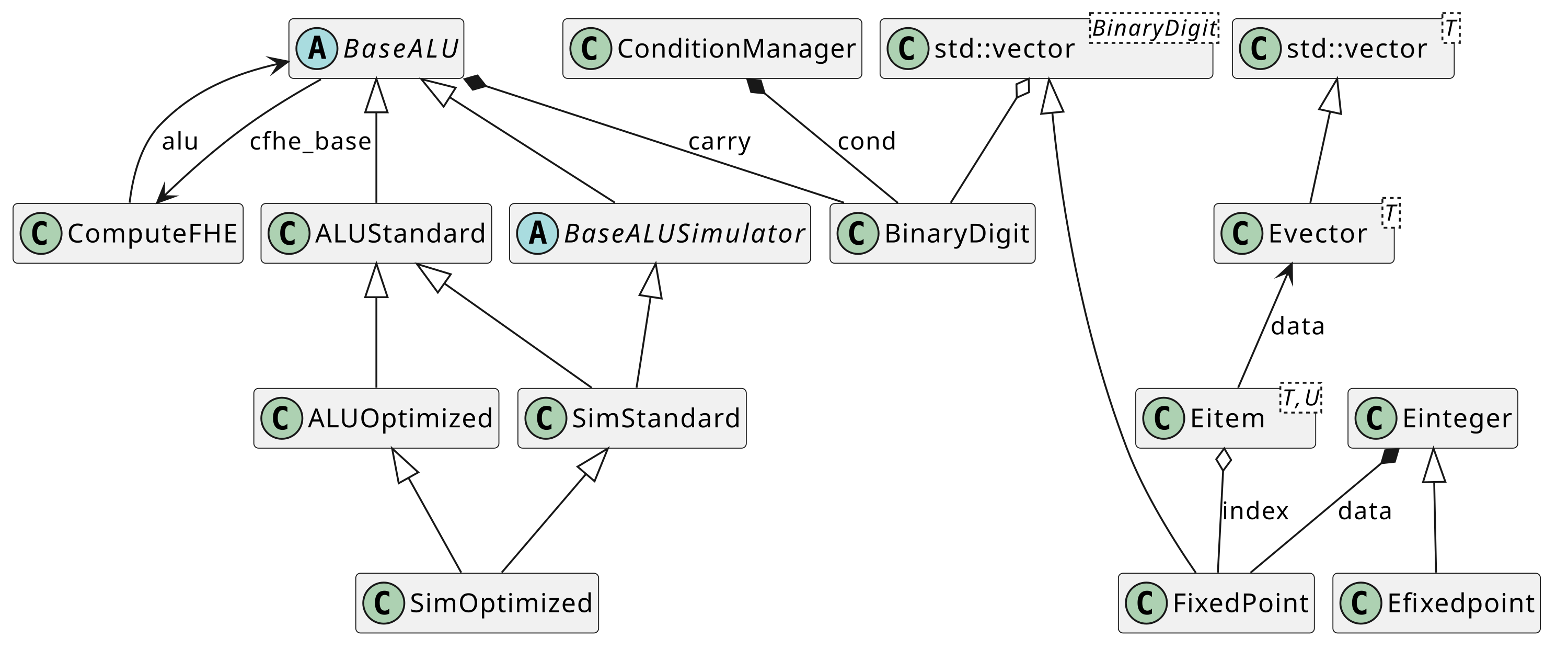}
    \caption{UML Diagram for ComputeFHE}
    \label{fig:uml}
    \end{figure}
    
    These modules and their corresponding submodules are described below:
    \begin{enumerate}
        \item Core modules:
        \begin{itemize}
            \item \textbf{ComputeFHE:} The main module of the library, responsible for initializing the TFHE cryptographic context with various security levels and ALU optimization configurations, as well as integrating the associated submodules.
            \item \textbf{BaseALU:} The base class that provides an interface between the low-level implementations of arithmetic and logical operations and the application programming interfaces (APIs). It enables the integration of different ALU architectures and designs.
            \item \textbf{StandardALU:} Baseline ALU implementation utilizing two-input logic gates.
            \item \textbf{OptimizedALU:} Optimized ALU implementation utilizing specialized three-input logic gates and more efficient algorithms.
            \item \textbf{FixedPoint:} Fundamental fixed-point structure for the low-level API, serving as a proxy object for both plaintext and encrypted integral data types.
        \end{itemize}
        
        \item C++ utility modules:
        \begin{itemize}
            \item \textbf{Condition Manager:} It provides necessary mechanisms, which implements the functionality of encrypted conditional statements and manages the states of encrypted variables accordingly.
            \item \textbf{Einteger:} It provides an integer data structure supporting various bit widths, with arithmetic and logical operators defined for encrypted computation.
            \item \textbf{Efixedpoint:} It provides a fixed-point data structure supporting configurable integer and fractional bit widths, with fixed-point arithmetic operators defined for encrypted computation.
            \item \textbf{Evector:} A module extending the standard vector structure to enable access to encrypted data using both plain and encrypted indices.
        \end{itemize}
        
        \item Simulator modules:
        \begin{itemize}
            \item \textbf{BaseALU Simulator:} The base class responsible for enabling the integration of different ALU simulators.
            \item \textbf{StandardALU Simulator:} A module that enables the \textbf{StandardALU} module to operate in simulation mode.
            \item \textbf{OptimizedALU Simulator:} A module that enables the \textbf{OptimizedALU} module to operate in simulation mode.\\
        \end{itemize}
    \end{enumerate}

    \newpage
    \item Software functionalities: ComputeFHE library has several functionalities that enable homomorphic data processing by adopting an imperative programming paradigm over encrypted data, as summarized below:
    \begin{itemize}
        \item The library provides template-based integral data types supporting arbitrary bit widths (up to 64 bits) (e.g., \texttt{EInt<n\_bits, is\_signed>}) and fixed-point data types parameterized by the programmer (e.g., \texttt{EFix<n\_bits, n\_fraction\_bits, is \_signed>}), as well as predefined C-like encrypted signed and unsigned integer types such as \texttt{Eint8} and \texttt{Euint32}. Variables of these types can be initialized and assigned in a C-like manner.
        \item Arithmetic and logical operations between encrypted data types (ciphertext -- ciphertext operations) are fully implemented within the library.
        \item Arithmetic and logical operations between encrypted and plain data types (ciphertext -- plaintext operations), which enable additional optimizations, are also implemented.
        \item Currently, the library supports both a baseline ALU (\texttt{ALU \_STANDARD}), implemented using conventional arithmetic computation methods \cite{tasel2025improved} based on basic two-input logic gates (e.g., AND, OR, XOR), and an optimized ALU (\texttt{ALU\_OPTIMIZED}) constructed using TFHE-friendly three-input logic gates such as Majority and MulAdd \cite{tasel2025improved,micciancio2021bootstrapping}.
        \item Oblivious array access is supported through the encrypted vector implementation (\texttt{Evector}).
        \item An implementation of encrypted conditional statements, provided through the \texttt{Eif} macro, is available to enable the modification of encrypted variables based on encrypted conditions.
        \item The library supports a client mode that enables automatic encryption through variable initialization and assignment, as well as automatic decryption during conversion to plaintext. Similarly, a server mode is provided to enable optimized operations on encrypted data where plaintexts are involved.
        \item The simulation mode emulates ALU operations efficiently for testing and debugging purposes, while reporting the underlying logical primitives together with the required bootstrapping operations to estimate real execution performance.\\
    \end{itemize}
\end{enumerate}

\item Illustrative examples\\

ComputeFHE comes with a selection of sample algorithms and examples. Among them is the \textit{quickstart} program\footnote{\scriptsize\url{https://github.com/fstasel/compute-fhe/blob/master/src/examples/quickstart.cpp}}, which introduces basics (i.e., initialization, variable declaration, encryption decisions) of ComputeFHE syntax to the user:\\

\begin{lstlisting}
#include <computefhe/ComputeFHE.h>
using namespace computefhe;
using namespace std;

int main() {
	// initialize with toy security and optimized ALU logic
	Init(CCPARAM_TOY, ALU_OPTIMIZED, true);

	// encrypt some values
	Euint8 a = 42;
	Euint8 b = 10;

	// perform homomorphic addition
	Euint8 sum = a + b;

	// use encrypted conditional branching
	Eif(sum > 50) {
		sum -= 5;
	} else {
		sum += 5;
	}

	// decrypt and print
	// (conversion to primitive types triggers decryption)
	cout << "Result: " << (uint32_t)sum << endl;

	// terminate cfhe-context
	Finalize();

	return 0;
}
\end{lstlisting}
\bigskip
The output of the \textit{quickstart} example is illustrated in Figure~\ref{fig:quickstart}. The standard implementation (Figure~\ref{fig:quickstarta}), which can be toggled by using \texttt{ALU\_STANDARD} as the 2nd parameter of the \texttt{Init()} function, requires 91 bootstrapping operations, whereas the optimized variant (Figure~\ref{fig:quickstartb}) reduces this count to 52.\\

\begin{figure}[bt]
{\phantomsubcaption\label{fig:quickstarta}\phantomsubcaption\label{fig:quickstartb}}
\centering
\includegraphics[width = \linewidth]{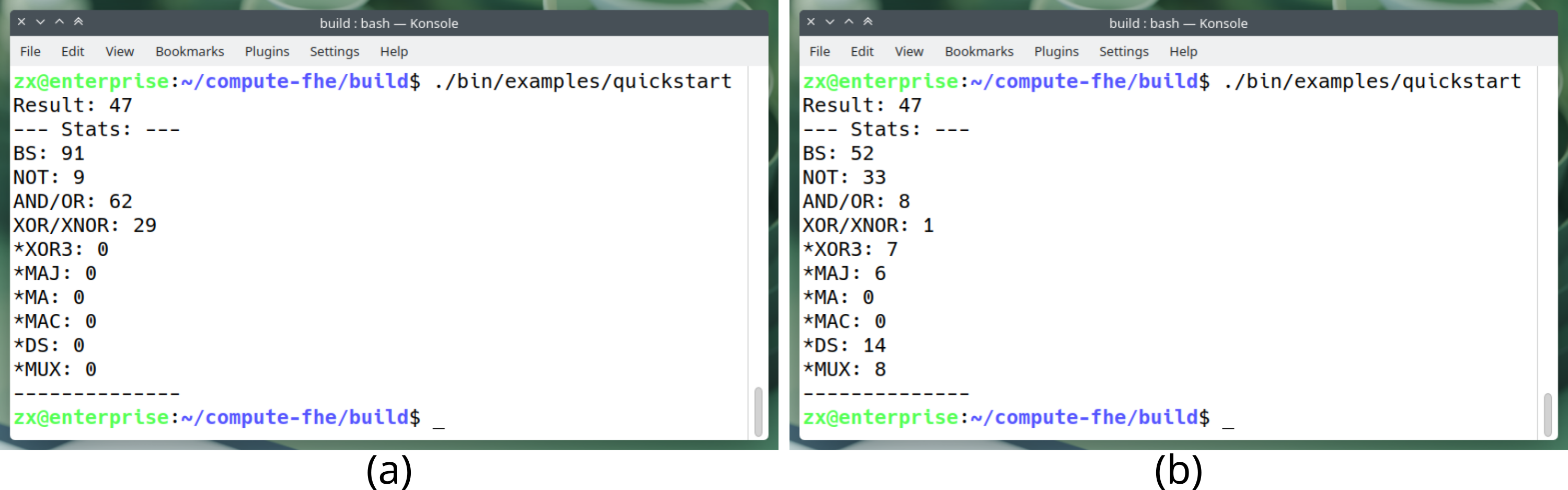}
\caption{Quickstart example. (a) Standard ALU implementation, (b) Optimized ALU implementation.}
\label{fig:quickstart}
\end{figure}

Similarly, the output for the sorting (Batcher odd-even mergesort algorithm) example\footnote{\scriptsize\url{https://github.com/fstasel/compute-fhe/blob/master/src/examples/sort.cpp}} is presented in Figure~\ref{fig:sort}. As with the previous example, the optimized version (Figure~\ref{fig:sortb}) performs significantly fewer bootstrapping operations (2,016) compared to the standard implementation (Figure~\ref{fig:sorta}), which requires 3,843 bootstrapping operations.\\

\begin{figure}[bt]
{\phantomsubcaption\label{fig:sorta}\phantomsubcaption\label{fig:sortb}}
\centering
\includegraphics[width = \linewidth]{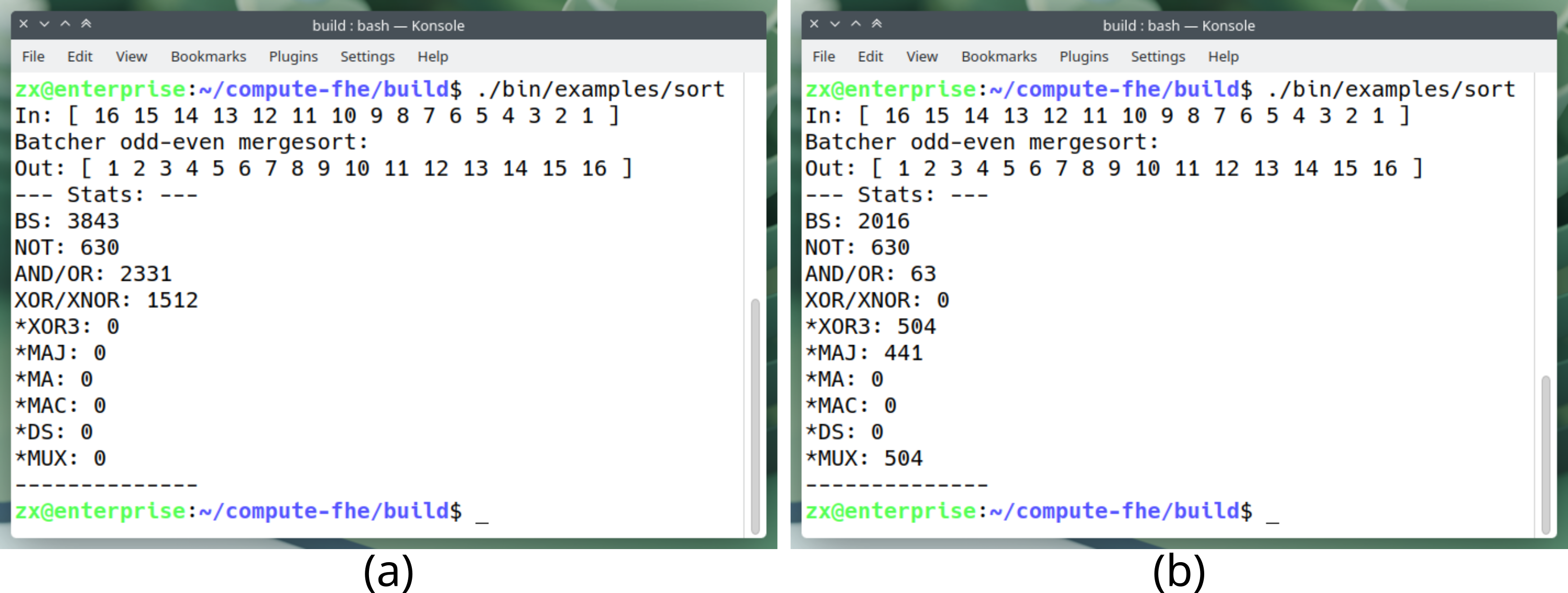}
\caption{Encrypted sort example. (a) Standard (b) Optimized}
\label{fig:sort}
\end{figure}

The \texttt{src/examples} directory of the ComputeFHE repository provides further implementations illustrating how to execute encrypted operations on several data types, such as integers, fixed-point numbers, character strings, and oblivious-indexed arrays.\\

\item Impact\\

Homomorphic encryption development typically requires the use of complex techniques and a substantial understanding of the underlying cryptographic mechanisms. In contrast, deep expertise in homomorphic encryption is not required for developers to begin using ComputeFHE. The library provides an intuitive, C-like programming interface that enables rapid development of applications operating on encrypted data. Encrypted variables can be created using encrypted counterparts of familiar C-style data types. Initialization and assignment operations automatically trigger encryption, while conversion back to plaintext automatically performs decryption. As a result, developers can perform arithmetic, logical, and comparison operations on encrypted variables in much the same way as they would on conventional C/C++ variables.\\

In addition to encrypted integer data types, the library also supports encrypted fixed-point data types for processing approximate fractional values. Although fixed-point representations do not provide the same flexibility as floating-point arithmetic, they offer significantly lower computational overhead, making them particularly attractive in the context of FHE. Furthermore, all data types provided by the library can be instantiated with application-specific bit widths. This capability is especially important in FHE applications, where minimizing operand sizes can substantially reduce computational cost and avoid unnecessary operations.\\

The library supports encrypted conditional statements (i.e., \texttt{Eif}), allowing nested \texttt{if} and \texttt{else} constructs based on encrypted Booleans. Internally, this functionality is realized by tracking variables affected by encrypted conditions and automatically resolving the final program state through multiplexing operations based on the outcome of the corresponding encrypted predicates. This allows developers to express conditional logic using familiar imperative programming constructs while preserving data confidentiality.\\

Another important feature of the library is support for oblivious array access. To achieve this, the standard C++ \texttt{vector} container has been extended to provide both read and write access to encrypted array elements using either plaintext or encrypted indices. This functionality enables the implementation of algorithms whose memory-access patterns must remain hidden from the service provider.\\

ComputeFHE utilizes OpenFHE as its backend framework. The fundamental homomorphic logic-gate implementations provided by OpenFHE form the basis of the standard ALU implementation. In addition, the specialized gates required for the optimized ALU design have been implemented using OpenFHE's low-level homomorphic primitives. Performance improvements provided by OpenFHE, including parallelized execution and hardware acceleration through the HEXL library, are naturally inherited by ComputeFHE and directly contribute to its overall efficiency.\\

ComputeFHE provides two alternative ALU implementations: \texttt{StandardALU} and \texttt{OptimizedALU}. The \texttt{StandardALU} serves as a baseline implementation based on conventional homomorphic arithmetic techniques. In contrast, \texttt{OptimizedALU} employs FHE-friendly specialized gates to perform arithmetic, logical, and shift operations more efficiently. These optimizations reduce the number of required bootstrapping operations and consequently improve performance. Furthermore, the error characteristics of the optimized implementations have been investigated and compared against standard logic-gate-based designs, and their feasibility has been demonstrated in \cite{tasel2025improved}. Compared with the standard implementation, the optimized version provides substantial execution-time improvements at the cost of only a modest increase in memory consumption. Performance comparisons for selected operations are presented in Table \ref{table:comparison}.\\

\begin{table}[bt]
\footnotesize
\centering
\begin{threeparttable}
    \begin{tabular}{r|c|c|c|c}
    & & \multicolumn{2}{c|}{\textbf{\# of BS Operations}} & \\
    \textbf{Operation}   & \textbf{\# of bits} & \textbf{ALU\_STD} & \textbf{ALU\_OPT} & \textbf{Gain} \\\hline\hline
    \textbf{Add / Sub}   &                     & 34                & 15                & 2.27×         \\
    \textbf{Mul}         &                     & 136               & 78                & 1.74×         \\
    \textbf{Div / Mod}   & \textbf{8-bit}      & 528               & 248               & 2.13×         \\
    \textbf{Equ / Inequ} &                     & 15                & 15                & 1×            \\
    \textbf{Cmp}         &                     & 29                & 8                 & 3.63×         \\\hline
    \textbf{Add / Sub}   &                     & 74                & 31                & 2.39×         \\
    \textbf{Mul}         &                     & 648               & 346               & 1.87×         \\
	\textbf{Div / Mod}   & \textbf{16-bit}     & 2328              & 1008              & 2.31×         \\
    \textbf{Equ / Inequ} &                     & 31                & 31                & 1×            \\
    \textbf{Cmp}         &                     & 61                & 16                & 3.81×         \\\hline
    \textbf{Add / Sub}   &                     & 154               & 63                & 2.60×         \\
    \textbf{Mul}         &                     & 2824              & 1458              & 1.94×         \\
    \textbf{Div / Mod}   & \textbf{32-bit}     & 9768              & 4064              & 2.40×         \\
    \textbf{Equ / Inequ} &                     & 63                & 63                & 1×            \\
    \textbf{Cmp}         &                     & 125               & 32                & 3.91×         \\\hline
    \textbf{Integer Sqrt}\tnote{a}& \textbf{64-bit}     & 9954     & 5706              & 1.74×         \\\hline
    \multicolumn{2}{r|}{\textbf{Oblivious array access (read)}}
                                               & 2.23 \tnote{b}    & 1.23 \tnote{b}    & 1.81×         \\
    \multicolumn{2}{r|}{\textbf{Oblivious array access (write)}}
                                               & 3.23 \tnote{b}    & 2.23 \tnote{b}    & 1.45×         \\
    \multicolumn{2}{r|}{\textbf{Batcher Odd-even Mergesort}}
                                               & 1.79 \tnote{c}    & 0.94 \tnote{c}    & 1.91×
    \end{tabular}
    \begin{tablenotes}
        \item[a] Binary restoring square root algorithm
        \item[b] Average \# of BS per (array size $\times$ bitsize)
        \item[c] Average \# of BS per ($nlog_2^2(n)$ $\times$ bitsize) where $n$ denotes the input size
    \end{tablenotes}
\end{threeparttable}
\caption{Number of bootstrapping operations required using standard and optimized ALU and corresponding gains across different homomorphic operations}
\label{table:comparison}
\end{table}

As shown in the table, the optimized ALU significantly reduces the number of required bootstrapping operations compared with the standard ALU implementation. One notable exception is the equality and inequality operators, for which the bootstrapping count remains unchanged, as the corresponding standard implementation is already near-optimal. As expected, the number of required bootstrapping operations increases with the bit width of the data type. However, the performance advantage of the optimized ALU also becomes more effective as the operand size grows, which results in greater relative savings for larger data types. It is also worth noting that oblivious array access benefits from the optimized ALU implementation; nevertheless, it still exhibits a computational complexity of $O(n)$. Consequently, although its execution cost can be reduced through ALU-level optimizations, oblivious array access remains computationally expensive and should be used sparingly to avoid unnecessary invocations whenever possible.\\

ComputeFHE further improves efficiency by executing ciphertext -- plaintext operations more efficiently than ciphertext -- ciphertext operations, thereby reducing the number of costly homomorphic computations and avoiding unnecessary bootstrapping procedures whenever possible.\\

\newpage
Although ComputeFHE includes optimized arithmetic and logical implementations, additional optimizations may still be possible at the algorithm-design level. Certain algorithms are inherently more FHE-friendly than others and may therefore achieve superior performance. Moreover, because different arithmetic operations incur different computational costs, avoiding unnecessary arithmetic computations, reformulating expressions using lighter-weight operations, or replacing arithmetic with logical operations when appropriate can significantly improve efficiency. To facilitate such low-level optimizations, ComputeFHE provides access to ALU intrinsics. These interfaces allow developers to invoke instruction-level operations directly and access internal ALU flags, such as the carry flag.\\

Furthermore, new instructions can be developed and integrated into the framework whenever a particular operation can be implemented more efficiently in TFHE than through conventional C/C++ constructs. For example, the specialized \texttt{SwapIf} instruction provides a more efficient implementation of conditional swapping than a conventional \texttt{if}-based swap operation.\\

One of the most significant features of ComputeFHE is its simulation mode. This mode enables execution simulation without requiring any modifications to the application code. It allows developers to estimate the homomorphic complexity of an algorithm, including its computational and memory requirements, without performing actual cryptographic computations. Consequently, the simulator serves as a valuable tool for debugging, testing, and early-stage performance analysis.\\

ComputeFHE can be installed conveniently through Docker, providing a fast and hassle-free deployment experience. On bare-metal systems, installation typically requires configuring a suitable GNU/Linux development environment, installing the necessary compiler toolchains and dependencies, and subsequently building and installing OpenFHE. Docker eliminates much of this setup complexity. Using the Dockerfile distributed with ComputeFHE, users can generate a fully configured ComputeFHE image with all dependencies preinstalled. Modern development environments such as Visual Studio Code can connect directly to this containerized environment through Docker extensions, which enables platform-independent software development using ComputeFHE.

Currently, ComputeFHE does not provide support for encrypted floating-point arithmetic. Although this limitation is partially mitigated through the availability of encrypted fixed-point data types, fixed-point arithmetic cannot fully match the flexibility and dynamic range offered by floating-point representations. In addition, oblivious array access remains computationally expensive and would benefit from further optimization. Consequently, developers are strongly encouraged to design algorithms that minimize unnecessary array accesses whenever possible.\\

\item Conclusion\\

ComputeFHE is an easy-to-use C++ library that enables fully homomorphic computation without requiring advanced expertise in FHE. The library provides implementations of encrypted integral and fixed-point data types together with the arithmetic and logical operations. Moreover, it includes encrypted conditional statements and oblivious array access mechanisms, which allow developers to design algorithms over encrypted data while maintaining a familiar programming model.\\

One of the distinguishing features of the library is its support for both conventional TFHE arithmetic implementations based on standard two-input logic gates and optimized ALU architectures built using FHE-friendly logic primitives. These optimized implementations significantly reduce the number of required bootstrapping operations and improve overall computational efficiency.\\

Another important feature is the simulation mode, which allows applications to be executed for testing and debugging purposes without requiring substantial computational resources or memory consumption and without any modifications to the original implementation. During execution, the simulator also provides statistical information about the underlying circuits, including estimates of the logical operations and bootstrapping costs that would be incurred during actual homomorphic execution.\\

Several research directions are planned as future work. These include the investigation of optimized encrypted floating-point representations, which are also of considerable interest in machine learning applications, as well as further optimizations for division operations, oblivious array access, and blind shift and rotation operations in which the shift amount remains encrypted. In addition, research will focus on the automatic detection of optimization opportunities within application code, including techniques for eliminating redundant recomputations and improving circuit efficiency.\\

Additional areas of interest include automatic detection and optimization of squaring operations, further optimization of encrypted conditional statements, and higher-level parallelization strategies aimed at improving CPU and GPU utilization during homomorphic computation.\\

Furthermore, depending on application requirements, future work will explore optimized implementations of encrypted algorithms in several application domains, including encrypted signal and image processing, privacy-preserving machine learning, and privacy-preserving steganography and steganalysis.\\

\end{enumerate}










\bibliographystyle{elsarticle-num} 
\bibliography{refs}

@article{doan2023survey,
  title={A survey on implementations of homomorphic encryption schemes: DT Van Thao et al.},
  author={Doan, Thi Van Thao and Messai, Mohamed-Lamine and Gavin, G{\'e}rald and Darmont, J{\'e}r{\^o}me},
  journal={The Journal of Supercomputing},
  volume={79},
  number={13},
  pages={15098--15139},
  year={2023},
  publisher={Springer},
  doi={10.1007/s11227-023-05233-z}
}

@article{tasel2025improved,
  title={Improved arithmetic efficiency in TFHE through gate-level optimizations: FS Ta{\c{s}}el, AN Saran},
  author={Ta{\c{s}}el, Faris Serdar and Saran, Ay{\c{s}}e Nurdan},
  journal={The Journal of Supercomputing},
  volume={81},
  number={18},
  pages={1633},
  year={2025},
  publisher={Springer},
  doi={10.1007/s11227-025-08107-8}
}

@inproceedings{cheon2018full,
  title={A full RNS variant of approximate homomorphic encryption},
  author={Cheon, Jung Hee and Han, Kyoohyung and Kim, Andrey and Kim, Miran and Song, Yongsoo},
  booktitle={International Conference on Selected Areas in Cryptography},
  pages={347--368},
  year={2018},
  organization={Springer},
  doi={10.1007/978-3-030-10970-7_16}
}

@inproceedings{chillotti2016faster,
  title={Faster fully homomorphic encryption: Bootstrapping in less than 0.1 seconds},
  author={Chillotti, Ilaria and Gama, Nicolas and Georgieva, Mariya and Izabachene, Malika},
  booktitle={international conference on the theory and application of cryptology and information security},
  pages={3--33},
  year={2016},
  organization={Springer},
  doi={10.1007/978-3-662-53887-6_1}
}

@inproceedings{chillotti2017faster,
  title={Faster packed homomorphic operations and efficient circuit bootstrapping for TFHE},
  author={Chillotti, Ilaria and Gama, Nicolas and Georgieva, Mariya and Izabach{\`e}ne, Malika},
  booktitle={International Conference on the Theory and Application of Cryptology and Information Security},
  pages={377--408},
  year={2017},
  organization={Springer},
  doi={10.1007/978-3-319-70694-8_14}
}

@Misc{TFHE-rs,
  title={{TFHE-rs: A Pure Rust Implementation of the TFHE Scheme for Boolean and Integer Arithmetics Over Encrypted Data}},
  author={Zama},
  year={2022},
  note={\url{https://github.com/zama-ai/tfhe-rs}},
}

@Misc{Concrete,
  title={{Concrete: TFHE Compiler that converts python programs into FHE equivalent}},
  author={Zama},
  year={2022},
  note={\url{https://github.com/zama-ai/concrete}},
}

@Misc{ConcreteML,
  title={Concrete {ML}: a Privacy-Preserving Machine Learning Library using Fully Homomorphic Encryption for Data Scientists},
  author={Zama},
  year={2022},
  note={\url{https://github.com/zama-ai/concrete-ml}},
}

@misc{IntelHEXL,
        author={Boemer, Fabian and Kim, Sejun and Seifu, Gelila and de Souza, Fillipe DM and Gopal, Vinodh and others},
        title = {{I}ntel {HEXL} (release 1.2)},
        howpublished = {\url{https://github.com/intel/hexl}},
        month = {september},
        year = {2021},
        key = {Intel HEXL}
    }

@article{gouert2025helm,
  title={Helm: Navigating homomorphic encryption through gates and lookup tables},
  author={Gouert, Charles and Mouris, Dimitris and Tsoutsos, Nektarios Georgios},
  journal={IEEE Transactions on Information Forensics and Security},
  year={2025},
  publisher={IEEE},
  doi={10.1109/TIFS.2025.3544066}
}

@article{chaturvedi2026bolt,
  title={BOLT: Bootstrapping-Aware Logic Resynthesis and Technology Mapping for Efficient TFHE Circuits},
  author={Chaturvedi, Bhuvnesh and Chatterjee, Ayantika and Chattopadhyay, Anupam and Mukhopadhyay, Debdeep},
  journal={Cryptology ePrint Archive},
  year={2026}
}

@article{bai2024research,
  title={Research on noise management technology for fully homomorphic encryption},
  author={Bai, Lifang and Bai, Lijuan and Li, Yongjun and Li, Zecun},
  journal={IEEE Access},
  volume={12},
  pages={135564--135576},
  year={2024},
  publisher={IEEE},
  doi={10.1109/ACCESS.2024.3461729}
}

@inproceedings{micciancio2021bootstrapping,
  title={Bootstrapping in FHEW-like cryptosystems},
  author={Micciancio, Daniele and Polyakov, Yuriy},
  booktitle={Proceedings of the 9th on Workshop on Encrypted Computing \& Applied Homomorphic Cryptography},
  pages={17--28},
  year={2021},
  doi = {10.1145/3474366.3486924}
}

@inproceedings{gentry2009fully,
  title={Fully homomorphic encryption using ideal lattices},
  author={Gentry, Craig},
  booktitle={Proceedings of the forty-first annual ACM symposium on Theory of computing},
  pages={169--178},
  year={2009},
  doi={10.1145/1536414.1536440}
}

@inproceedings{chillotti2021improved,
  title={Improved programmable bootstrapping with larger precision and efficient arithmetic circuits for TFHE},
  author={Chillotti, Ilaria and Ligier, Damien and Orfila, Jean-Baptiste and Tap, Samuel},
  booktitle={International Conference on the Theory and Application of Cryptology and Information Security},
  pages={670--699},
  year={2021},
  organization={Springer},
  doi={10.1007/978-3-030-92078-4_23}
}

@inproceedings{xiang2023fast,
  title={Fast blind rotation for bootstrapping FHEs},
  author={Xiang, Binwu and Zhang, Jiang and Deng, Yi and Dai, Yiran and Feng, Dengguo},
  booktitle={Annual International Cryptology Conference},
  pages={3--36},
  year={2023},
  organization={Springer},
  doi={10.1007/978-3-031-38551-3_1}
}

@inproceedings{bergerat2025accelerating,
  title={Accelerating tfhe with sorted bootstrapping techniques},
  author={Bergerat, Loris and Orfila, Jean-Baptiste and Roux-Langlois, Adeline and Tap, Samuel},
  booktitle={International Conference on the Theory and Application of Cryptology and Information Security},
  pages={68--100},
  year={2025},
  organization={Springer},
  doi={10.1007/978-981-95-5122-4_3}
}



\end{document}